\documentstyle{aipproc}

\begin{document}

\title{Can Quantum Cosmology Give Observational\\
Consequences of Many-Worlds Quantum Theory?}
\author{Don N. Page}
\address{CIAR Cosmology Program,
Institute for Theoretical Physics\\
Department of Physics, University of Alberta\\
Edmonton, Alberta, Canada T6G 2J1}

%\lefthead{LEFT head}
%\righthead{RIGHT head}
\maketitle

\begin{abstract}

	Although many people have thought
that the difference between the Copenhagen
and many-worlds versions of quantum theory
was merely metaphysical,
quantum cosmology may allow us to make
a physical test to distinguish between them empirically.
The difference between the two versions
shows up when the various components
of the wavefunction have different
numbers of observers and observations.
In the Copenhagen version,
a random observation is selected from
the sample within the component that
is selected by wavefunction collapse,
but in the many-worlds version,
a random observation is selected
from those in all components.
Because of the difference in the samples,
probable observations in one version
can be very improbable in the other version.

\end{abstract}

\section*{Introduction}

	Ever since Hugh Everett III formulated
his many-worlds alternative
\cite{everett57,dewitt73}
to the Copenhagen version of quantum theory,
there has been considerable discussion
of its merits.
Many people, including some of the original
supporters of the many-worlds version,
have expressed the opinion that
the many-worlds version
is empirically indistinguishable
from the Copenhagen version,
so that the difference is merely metaphysical.

	For example, in the first wide popularization
of the many-worlds or Everett-Wheeler-Graham (EWG)
version of quantum theory,
Everett's bulldog Bryce DeWitt stated
\cite{dewitt70},
``Clearly the EWG view of quantum mechanics leads to experimental predictions identical with those
of the Copenhagen view.''

	Everett's Ph.D. supervisor John Wheeler,
who initially supported the many-worlds version
\cite{wheeler57},
has recently summarized it as follows
\cite{wheeler98}:
``Does it offer any new insights?
Does it predict outcomes of experiments that differ
from outcomes predicted in conventional quantum theory?
The answer to the first question is emphatically yes.
The answer to the second question is emphatically no.''

	Roland Omn\`{e}s, though never a supporter
of the many-worlds version to my knowledge,
has thought about it deeply and concluded
\cite{omnes94},
``If quantum mechanics were absolutely true
and Everett were right,
no experiment would be able to confirm or reject it. \ldots
It is not science because no experiment
can show it to be wrong.''

	However, David Deutsch has argued
\cite{deutsch85}
that the many-worlds version of quantum theory
would be confirmed if an observer could ``split''
into two copies which make different observations,
remember that they observed but not what they observed,
and then are rejoined coherently.
Although there are conceptual loopholes
(such as claiming after the experiment
that the observer's memory of having made
a definite observation is merely a false memory),
I believe this argument is fairly strong evidence
that the difference between the Copenhagen
and many-worlds versions of quantum theory
is, in principle at least, a matter that could
be experimentally tested.
Nevertheless, this proposed test appears to
be technically very difficult.

	Because of the difficulties
of Deutsch's proposed experiment,
here I wish to raise the possibility
that quantum cosmology might in principle
lead to empirical distinctions between
the Copenhagen and the many-worlds versions
of quantum theory.

	By the Copenhagen version,
I essentially mean what I might
more accurately call a single-history version,
in which quantum theory gives probabilities
for various alternative sequences of events,
but only one sequence actually occurs.
Each such alternative sequence
might be called a ``history'' or a ``world.''

	In the many-worlds version, in contrast,
all of the possible histories or worlds
with nonzero quantum probabilities
actually occur, with the quantum probabilities
being not probabilities for the histories to be actualized
(since all are), but instead essentially measures for the magnitude of the existence of the various histories.

\section*{Consequences of Different Numbers of
Observers}

	There can be significant differences in
typical observations if the number of observers
varies greatly from ``world'' to ``world.''
Consider the following toy models:

	Quantum Cosmology Model I\\
World 1:  Observers; measure or probability $10^{-100}$\\
World 2:  No observers; measure or probability $1-10^{-100}$

	In a single-history version of this Model I,
World 1 is very improbable to occur at all,
so any observation would be strong evidence
against the single-history version.
In a many-worlds version, World 1 does occur,
so observations are not evidence against that theory.

	Quantum Cosmology Model II\\
World A:  $10^{10}$ observers during collapse;
measure $1-10^{-30}$\\
World B:  $10^{90}$ observers during expansion;
measure $10^{-30}$

	In a single-history version of Model II,
World B is very improbable, so a random observation
should expect to see a collapsing universe,
Hubble constant $H < 0$,
and the probability that $H > 0$
is observed is only $10^{-30}$.

	In contrast, in a many-worlds version of Model II,
all of the observations occur,
with measures presumably given by something like
the expectation values of positive operators
each associated with a corresponding observation
\cite{page95,page96}.
I shall assume that the observers are sufficiently
similar that the total measure
of a certain set of observations in a certain world
(e.g., of whether the universe is expanding)
is roughly proportional to the total number
of observers in that world who make the observation,
multiplied by the quantum measure of that world.
I shall also assume that the fraction of observers
who do observe whether the universe is expanding
or contracting is the same in both World A and World B.

	Then the total measure for World A observations
of a collapsing universe is roughly proportional
to the $10^{10}$ observers times
the quantum measure of nearly unity for that world,
or $10^{10}$, whereas the total measure for
World B observations of an expanding universe is
roughly proportional to $10^{90}$ observers
of that world times the quantum measure of $10^{-30}$
for that world, or $10^{60}$.
Thus a random observation chosen from the
sample of all existing observations
in the many-worlds version
is about $10^{50}$ times more likely
to be from World B, seeing $H > 0$, than
it is to be from World A, seeing $H < 0$,
a situation qualitatively the reverse
of the relative probabilities in a single-history
version, such as the Copenhagen version
of quantum theory.

	Now if one accepted the basic quantum measures
of the two worlds in Quantum Cosmology Model II
but was not very certain whether a single-history
or a many-worlds version of quantum theory were
correct, then if one made an observation of
whether the universe were expanding or contracting,
it would give strong evidence as to which version
is correct.

	One way to explain the difference between
sampling a random observation in single-history
versus many-worlds quantum theories
is with lottery tickets.  Suppose that we have
a quantum cosmological model with the
following two worlds:\\
World 1:  $N_1$ observers;
quantum measure or probability $p_1$\\
World 2:  $N_2$ observers;
quantum measure or probability $p_2$

	The single-history version of quantum theory
is like assigning lottery tickets to World 1 and World 2
in the ratio $p_1: p_2$.
Then a lottery ticket is chosen at random to select
which world, and its observers, exist.

	The many-worlds version of quantum theory
is like assigning lottery tickets to each observer
in World 1 and 2 with ratio $p_1: p_2$,
so that the ratio of the total number of lottery tickets
in world 1 to that in world 2 is $N_1 p_1: N_2 p_2$.
All the observers exist, but with different measures
for their reality, analogous to holding different
numbers of lottery tickets.
Choosing a measure-weighted observer
(or, better, observation) at random
is analogous to choosing a lottery ticket at random.
The choice really is not made (since all observations
really exist in the many-worlds version),
but for saying which observations are typical,
is is helpful to imagine their being chosen randomly.

\section*{Preliminary Evidence from Hartle-Hawking}

	We cannot yet calculate probabilities
for our observations from an accepted model
of the quantum cosmology quantum measures,
so we cannot yet perform a definitive test
of whether the single-history or the many-worlds
version of quantum theory is correct.
However, we can examine some highly speculative
preliminary suggestions from the Hartle-Hawking
`no-boundary' proposal
\cite{hawking82,hartle83,hawking84}
applied to a $k=+1$ Friedmann-Robertson-Walker model
with a minimally coupled massive scalar field
(potential ${1\over 2}m^2\phi^2$).

	In this minisuperspace model,
an approximation of the stationary
phase approximation for the path integral
in which the scalar field starts at a value $\phi_i$
large compared with the Planck value (unity here)
leads to the universe nucleating with initial size
\begin{equation}
a_i^2 = {3 \over 4 \pi m^2 \phi_i^2} = {p\over \pi}
\end{equation}
and quantum measure roughly proportional to
\begin{equation}
e^{-2S_E} \approx e^{\pi a_i^2} = e^p
\end{equation}
with $p \equiv \pi a_i^2$.
Observations suggest $m \sim 10^{-6}$
\cite{linde90}.

	This is actually a measure density,
and it is not clear what the prefactor should be.
One simple choice is $dp = 2\pi a_i d a_1$.
The resulting measure would diverge if
integrated to $p = \infty$ or $a_i = \infty$,
but this would correspond to $\phi_i = 0$,
where the approximation is invalid.
To get an inflationary solution,
one needs $\phi_i > \phi_{\rm min} \sim 1$, so
$a_i < a_m = \sqrt{3/4\pi}/(m \phi_{\rm min}) \sim 1/m$
or $p < p_m = 3/(4 m^2 \phi_i^2) \sim 1/m^2$.
Cut off the measure density there and normalize it,
so we get the simple idealization
\begin{equation}
P(p < p') \approx {e^{p'} - 1 \over e^{p_m} - 1}
\approx e^{-p_m}(e^{p'}-1)
\end{equation}
for $p' \equiv \pi a_i^2 < p_m \equiv \pi a_m^2 \sim 1/m^2
\sim 10^{12} \gg 1$.

	After the universe nucleates,
it undergoes slow-roll inflation with
$\phi$ decreasing from $\phi_i$ to $\phi_e \sim 1$
and the volume increasing to
\begin{equation}
V_e = V_i\left({a_e\over a_i}\right)^3
\approx {\sqrt{27\pi}\over 4 m^3 \phi_i^3}
e^{6\pi(\phi_i^2 - \phi_e^2)}
= 2\sqrt{\pi}e^{-6\pi\phi_e^2}p^{3/2}\exp{4.5\pi\over m^2 p}
\sim p^{3/2}\exp{4.5\pi\over m^2 p},
\end{equation}
which implies that for $m^3 V_e \gg 1$,
\begin{equation}
p \sim {4.5\pi/m^2 \over \ln{(m^3V_e)}
+ 1.5 \ln{\ln{(m^3V_e)}}}.
\end{equation}
The entropy density after reheating is
\begin{equation}
s_e \sim T_e^3 \sim \rho_e^{3/4}
\sim (m^2\phi_e^2)^{3/4}
\sim m^{3/2} \sim 10^{-9}.
\end{equation}
By comparison, the entropy density of radiation today is
\begin{equation}
s_0 \approx {86 \pi^2 \over 165} T_0^3
\approx 1.22 \times 10^{-95}.
\end{equation}
Assuming essentially adiabatic expansion after reheating,
one gets that the volume of the universe today is
\begin{equation}
V_0 \approx {s_e \over s_0} V_e \sim 10^{95} m^{3/2} V_e
\sim 10^{95} m^{3/2} p^{3/2} \exp{4.5\pi\over m^2 p}
\sim 10^{86} p^{3/2} e^{1.4\times 10^{13}/p}.
\end{equation}

	Now to get something analogous to
Quantum Cosmology Model II above,
we need to consider what values of
$p$ give observers mainly seeing the universe
either contracting or expanding,
and how many observers are produced
as a function of $p$.

	Let us make the crude assumption that
observers require a universe of an age
at least of the order of $10^{60}$,
a tenth of the age of our actual universe,
and hence a volume of the order of $10^{181}$,
in order for suitable habitats to have evolved
(e.g., planets around stars).
This would give a lower limit on the volume
at the end of inflation of about
\begin{equation}
V_e {\ \lower-1.2pt\vbox{\hbox{\rlap{$>$}
\lower5pt\vbox{\hbox{$\sim$}}}}\ }
10^{86} m^{-3/2} \sim 10^{95}.
\end{equation}
Inserting this back into the approximate relation
between $V_e$ and $p$ gives
\begin{equation}
p < p_{\rm max} \sim
{4.5\pi/m^2 \over \ln{(10^{77})} + 1.5 \ln{\ln{(10^{77})}}}
\sim {4.5\pi/m^2 \over 185} \sim 7.6\times 10^{10}
\end{equation}
as the crude condition for the existence of observers.

	However, if $p$ is sufficiently near this upper
limit $p_{\rm max}$ for the existence of observers,
then the universe will just barely last long enough
for them, and they will mostly exist near the end
of the lifetime of the universe, when it is collapsing.
For most observers to see the universe expanding,
$V_e$ must be sufficiently larger that the lifetime
of the universe is long enough for most observers
to exist while the universe is still expanding.
If the present age of the universe is a typical time
for observers, then one might estimate that the universe
must still be expanding at an age of roughly $10^{61}$
for most observers to see the universe expanding,
and hence for it to have a volume of at least
of the order of $10^{184}$ then.
This leads us to
$V_e {\ \lower-1.2pt\vbox{\hbox{\rlap{$>$}
\lower5pt\vbox{\hbox{$\sim$}}}}\ }
10^{89} m^{-3/2} \sim 10^{98}$
and
\begin{equation}
p < p_{\rm exp} \sim
{4.5\pi/m^2 \over \ln{(10^{80})} + 1.5 \ln{\ln{(10^{80})}}}
\sim {4.5\pi/m^2 \over 192} \sim 7.4\times 10^{10}
\end{equation}
as the crude condition for most observers
to see the universe expanding.

	In other words, in the Hartle-Hawking
minisuperspace model under consideration,
if $0 < p < p_{\rm exp}\sim 7.4\times 10^{10}$,
observers will exist and will mostly see the universe
expanding;
if $p_{\rm exp} < p < p_{\rm max} \sim 7.6\times 10^{10}$,
observers will exist but will mostly see the universe
contracting; and if $p_{\rm max} < p$,
essentially no observers will exist.

	First, consider a Copenhagen or other
single-history version of this
quantum minisuperspace model,
in which the wavefunction collapses
to give a single macroscopic history or world,
a classical Friedmann-Robertson-Walker
universe characterized by $\phi_i$, $a_i$, or $p$.

	Using the results above,
the probability for the wavefunction to collapse
to classical universe that lasts long enough for observers is
\begin{equation}
P({\rm observers})
\approx {e^{p_{\rm max}} - 1 \over e^{p_m} - 1}
\approx e^{p_{\rm max} - p_m}
%\sim e^{-10^{12}}e^{7.6\times 10^{10}}
\sim 10^{-401000000000},
\end{equation}
and the probability for it to have observers
that mostly see the universe expanding is
\begin{equation}
P({\rm observers\ seeing\ expansion})
%\approx {e^{p_{\rm exp}} - 1 \over e^{p_m} - 1}
\approx e^{p_{\rm exp} - p_m}
%\sim e^{-10^{12}}e^{7.4\times 10^{10}}
\sim 10^{-402000000000},
\end{equation}
both of which are utterly tiny.

	Therefore, unless one had an uncertainty
less than roughly $e^{-0.924\times 10^{12}}\sim 10^{-40100000000}$ that this single-history model
was correct, the evidence that observers exist
%(``I observe, ergo I am an observer.'')
would be overwhelming evidence against it.

	Even if one somehow claimed that observers
were necessary (i.e., that the wavefunction could
not collapse to a world with no observers),
the conditional probability of a world with
observers mostly seeing the universe expand, given
the condition that observers exist, is only
\begin{equation}
P({\rm observers\ seeing\ expansion}|{\rm observers\ exist})
\approx e^{p_{\rm exp} - p_{\rm max}}
%\sim e^{-0.2\times 10^{10}}
\sim 10^{-1000000000}.
\end{equation}
Thus the observation that the universe is expanding
would be strong evidence against the single-history
version of this model.

	On the other hand, if one takes a many-worlds
version of this Hartle-Hawking minisuperspace model,
all components of the wavefunction exist that have
positive measure, no matter how small, so observers
will exist in a generalization of the model that
allows sufficient structure for observers.
Therefore, the existence of observers would not
be evidence against a many-worlds version
of a model sufficiently general to allow
observers within at least some components of the
wavefunction.

	However, we can still ask for the relative
probabilities of observations that the universe
is contracting or expanding.
In a many-worlds version, this will be roughly
proportional to the bare quantum probability
for each world multiplied by the the number
of observers for that world.
The unnormalized bare quantum probability
was given above as $dP \approx e^p dp$
for $0 < p < p_m \sim 1/m^2 \sim 10^{12}$.
For $0 < p < p_{\rm max}$, observers exist,
and it is reasonable to assume that the number of them
is very roughly proportional to the volume $V_0$
of the universe when the entropy density is
roughly the value of $10^{-95}$ that we observe,
assuming that our observations of this quantity
are typical.
Then in this range of $p$, we get an
unnormalized observational
probability density roughly proportional to
\begin{equation}
dP_{\rm obs} \approx V_0 dP \approx V_0 e^p dp
\sim 10^{95} m^{3/2} p^{3/2}
\exp{\left({4.5\pi\over m^2 p} + p\right)}dp.
\end{equation}

	The integral of this over $0 < p < p_{\rm max}$
diverges for $p\rightarrow 0$ ($V_0\rightarrow \infty$),
because of the divergence in the number of observers
there, so effectively all of the observational probability
occurs at that limit (infinitely large universes
with infinitely many observers, presumably
almost all seeing the universe expanding,
since the stars would have all burned out in the
infinite time it takes the infinitely large universe
to recollapse.)

	Therefore, in a many-worlds version of
this Hartle-Hawking quantum cosmological model,
one would presumably expect with very nearly unit
probability that a random observation would see the
universe expanding, the opposite of one's expectation
for a single-history version of the same model.
Thus if one accepted the basic model
and allowed at least some reasonable uncertainty
as to whether a single-history or a many-worlds
version of the model is correct (before considering
the evidence of the sign of the Hubble constant),
then one's observation of whether the universe is
expanding or contracting would give very strong evidence
in support of either the many-worlds or the single-history
version respectively.
This is very similar qualitatively to the toy
Quantum Cosmology Model II discussed above,
except here the quantitative differences are
even grossly more severe.

	Of course, this preliminary evidence from
a particular implementation of the Hartle-Hawking
no-boundary proposal is highly speculative
and is meant to be mainly illustrative,
because of the many uncertainties of the model.

\section*{Conclusions}

	If the amount of observations
(roughly, the number of observers)
varies for different wavefunction components,
then observation probabilities
depend on whether only one component
occurs in actuality (a single-history version
of quantum theory, where observations
truly are made only in one history, world, or
component of the wavefunction),
or whether many do (a many-worlds version,
where observations truly are made
in many histories, worlds, or components
of the wavefunction).
In particular, if components with relatively few observations
dominate the quantum amplitude,
but other components with testably different
observations dominate the expectation value
of the number of observations,
which observations are most probable varies
between single-history and many-worlds quantum theories.

	The Hartle-Hawking wavefunction might allow
a test from the observed expansion of the universe,
but as of now it is highly speculative whether it is correct
and what relative probabilities it would give
for observing the universe expanding.

	I acknowledge helpful discussions
with Meher Antia, Jerry Finkelstein, Valeri Frolov,
Jim Hartle, Jacques Mallah, and William Unruh.
This research was supported in part by
the Natural Sciences and Engineering Research
Council of Canada.

	A shorter version of this paper has been
circulated
\cite{page99}
and has been reported on in the lay literature
\cite{antia99}.


\begin{references}

\bibitem{everett57}
Everett, H. III, {\it Rev.\ Mod.\ Phys.}\ {\bf 29}, 454 (1957).

\bibitem{dewitt73}
DeWitt, B. S., and Graham, N., eds.,
{\it The Many-Worlds Interpretation
of Quantum Mechanics}, Princeton: 
Princeton University Press, 1973.

\bibitem{dewitt70}
DeWitt, B. S., {\it Physics Today}\ {\bf 23},
September 1970, p. 30.

\bibitem{wheeler57}
Wheeler, J. A., {\it Rev.\ Mod.\ Phys.}\ {\bf 29}, 463 (1957).

\bibitem{wheeler98}
Ford, K. W., and Wheeler, J. A.,
{\it Geons, Black Holes, and Quantum Foam:
A Life in Physics},
New York:  W. W. Norton, 1998, p. 270.

\bibitem{omnes94}
Omn\`{e}s, R.,
{\it The Interpretation of Quantum Mechanics},
Princeton:  Princeton University Press, 1973, pp. 327, 345.

\bibitem{deutsch85}
Deutsch, D., {\it Int.\ J.\ Theor.\ Phys.}\ {\bf 24}, 1 (1985).

\bibitem{page95}
Page, D. N., "Sensible Quantum Mechanics:  Are Only
Perceptions Probabilistic?" (University of Alberta report
Alberta-Thy-05-95, June 7, 1995), quant-ph/9506010.

\bibitem{page96}
Page, D. N., {\it Int.\ J.\ Mod.\ Phys.}\ {\bf D5}, 583 (1996).

\bibitem{hawking82}
Hawking, S. W., in {\it Astrophysical Cosmology:
Proceedings of the Study Week on Cosmology and Fundamental Physics},
edited by Br\"{u}ck, H. A., Coyne, G. V. and Longair, M. S.,
Vatican:  Pontificiae Academiae Scientiarum Scripta Varia,
1982, pp. 563-574.

\bibitem{hartle83}
Hartle, J. B., and Hawking, S. W., 
{\it Phys.\ Rev.}\ {\bf D28}, 2960 (1983).

\bibitem{hawking84}
Hawking, S. W., {\it Nucl.\ Phys.}\ {\bf B239}, 257 (1984).

\bibitem{linde90}
Linde, A., {\it Particle Physics and Inflationary Cosmology},
Chur:  Harwood, 1990.

\bibitem{page99}
Page, D. N., "Observational Consequences
of Many-Worlds Quantum Theory"
(University of Alberta report
Alberta-Thy-04-99, May 3, 1999), quant-ph/9904004.

\bibitem{antia99}
Antia, M., {\it The Economist}, May 22, 1999, p. 145.

\end{references}
\end{document}